# Suppression of acoustic emission during superelastic tensile cycling of polycrystalline Ni$_{50.4}$Ti$_{49.6}$

Guillaume F. Nataf[a,*], Michela Romanini[b], Eduard Vives[b], Borut Žužek[c], Antoni Planes[b], Jaka Tušek[d,†], and Xavier Moya[a,‡]

[a]Department of Materials Science, University of Cambridge, 27 Charles Babbage Road, Cambridge, CB3 0FS, United Kingdom

[b]Departament de Física de la Matèria Condensada, Facultat de Física, Universitat de Barcelona, Martí i Franquès 1, 08028 Barcelona, Catalonia, Spain

[c]Institute of Metals and Technology, Lepi Pot 11, 1000 Ljubljana, Slovenia

[d]Faculty of Mechanical Engineering, University of Ljubljana, Aškerčeva 6, 1000 Ljubljana, Slovenia

*Corresponding author: gn283@cam.ac.uk, †Corresponding author: Jaka.Tusek@fs.uni-lj.si

‡Corresponding author: xm212@cam.ac.uk

We investigate acoustic emission (AE) that arises during the martensitic transition in a polycrystalline specimen of the prototypical superelastic/elastocaloric alloy Ni$_{50.4}$Ti$_{49.6}$ (at. %) driven using tensile strain. We use two independent AE sensors in order to locate AE events, and focus on contributions to the AE that arise away from the grips of the mechanical testing machine. Significant AE activity is present during the first mechanical loading primarily due to nucleation and growth of wide Lüders-like bands during the forward martensitic transition (imaged using visible light and infrared (IR) radiation) that lead to persistent changes in intergranular interactions. AE activity is suppressed during the subsequent reverse martensitic transition on unloading, and in successive loading/unloading cycles, for which the Lüders-like bands narrow and modify intergranular interactions to much less extent. After the first loading, we find that the AE activity associated with the martensitic transition is weak, and we suggest that this is because the elastic anisotropy and strain incompatibility in Ni-Ti are low. We also find that the AE activity becomes weaker on mechanically cycling due to increased retained martensite.

# I. INTRODUCTION

Ni-Ti alloys that are slightly rich in Ni content display thermally driven structural phase transitions near and below room temperature between a high-symmetry austenitic phase that is body-centred cubic (B2, space group $Pm\bar{3}m$), and a low-symmetry martensitic phase that is monoclinic (B19', space group $P2_1/m$), often via an intermediate martensitic phase that is rhombohedral (R, space group $P3$ or $P\bar{3}$) [1,2]. Instead, when being driven using stress or strain, these so-called martensitic transitions yield superelastic effects [2–4] and elastocaloric effects [5–8]. These functional properties make Ni-rich Ni-Ti alloys attractive for a number of industries, e.g. for applications in medical instrumentation and implants [4], actuators [3,4], or in refrigerators and air-conditioners [9–17].

For many such applications, Ni-Ti alloys are arguably preferred in polycrystalline form due to the relative ease of production and machining when compared to single crystals [18]. However, polycrystallinity introduces complexity to the thermomechanical properties of these alloys, and plays an important role in their fatigue life during prolonged operation [19–22]. For example, tensile-stress-driven transition fronts propagate across grains via nucleation and growth of localised macroscopic shear bands known as Lüders-like bands [7,23–27], whose complex hierarchical architecture is not yet fully understood despite extensive experimental and modelling work [28–30]. Moreover, x-ray diffractometry [31], transmission electron microscopy [32], and modelling [33] have recently revealed that a number of interconnected phenomena need to be considered simultaneously when understanding fatigue, namely retained martensite, residual strain due to gradual redistribution of internal stresses (which result in decreased transformation stresses and increased plastic deformation), and accumulation of defects at grain boundaries [31–33].

Here we use acoustic emission (AE) to investigate tensile-strain-driven B2-B19' transitions in a polycrystalline specimen of the superelastic/elastocaloric alloy $Ni_{50.4}Ti_{49.6}$. AE is a phenomenon by which alloys that display structural phase transitions can rapidly dissipate the accumulated elastic energy that is associated with differences in the lattice parameters of the transforming phases [34–37]. However, AE can also arise from irreversible dissipative processes within the material, such as dislocation motion, internal friction and fracture [38–40], and from irreversible dissipative processes caused by the grips of machines during mechanical testing, such as surface friction and plastic deformation [41–43]. AE from structural phase transitions normally does not occur smoothly but through avalanches, due to interactions between transition fronts and structural imperfections, such as dislocations, or chemical imperfections, such as local fluctuations in composition and impurities [37,43,44].

Near-equiatomic Ni-Ti polycrystalline alloys have been previously investigated using AE [45–47], but under a limited number of mechanical loading/unloading cycles, and without location of AE, which permits eliminating extrinsic contributions that arise near the grips of the mechanical testing machine. Dunand-Châtellet and Moumni assigned their total AE recorded during mechanical cycling to dislocation formation and micro-cracking [45], whereas Pieczyska *et al.* assigned their total AE recorded during one single cycle to the B2-B19' transition [46], despite AE being detected before the plateau in the stress-strain curve that marks the occurrence of the martensitic transition [2,48–50], suggesting that this AE could arise from the mechanical testing machine. In contrast, we use two independent sensors to locate and analyse AE that arises away from the grips of the mechanical testing machine, and we drive multiple times the martensitic transition using strain because irreversible energy dissipation is larger when driving the transition using stress [51]. Moreover, we record simultaneously optical and infrared images in order to compare AE data with nucleation and propagation of transition fronts. Our main findings are (i) that the AE activity is only strong during the first loading, due to persistent changes in grain configuration and dislocation density associated with the nucleation and propagation of Lüders-like bands that are wide, (ii) that the AE activity on subsequently driving the reverse transition is weak, which we attribute to the existence of two soft deformation modes for the transition, and (iii) that the AE activity becomes weaker with mechanical cycling due to the gradual increase in the amount of retained martensite.

## II. MATERIALS AND METHODS

### A. Samples

0.2-mm-thick $Ni_{50.4}Ti_{49.6}$ polycrystalline cold-rolled sheets, with average grain size ~45 μm, and front-surface roughness of ~0.5 μm, were purchased from Memry Corporation. The composition of the alloy was determined by averaging values obtained via energy dispersive x-ray analysis [$Ni_{50.3}Ti_{49.7}$ (at. %), with uncertainty of ~0.2 at. %] and x-ray fluorescence [$Ni_{50.5}Ti_{49.5}$ (at. %), with uncertainty of ~0.3 at. %]. Transition temperatures were determined via differential scanning calorimetry, with austenitic finish temperature ~290 K (Supplementary Note 1, [52]). Calorimetric measurements also revealed a thermally driven two-step martensitic transition B2 ↔ R ↔ B19'.

The cold-rolled sheets were cut along the rolling direction into dog-bone shapes that are suitable for mechanical testing [Figure 1(a)], using a waterjet cutter (samples SP1-3) and a wire electrical discharge machine (sample SP4). Waterjet cutting resulted in lateral-surface roughness of ~26 μm, whereas electrical discharge machining resulted in smoother lateral-surface roughness of ~9 μm (estimated using optical microscopy,

Supplementary Note 2, [52]). For samples SP1-4, the initial gauge length *L* was 80 mm, and the initial gauge width *W* was 10 mm [Figure 1(a)].

## B. Experimental methods

Mechanical testing was performed using a Zwick/Roell Z005 tensile testing machine (class ISO7500/1), equipped with two grips, and a 5kN XForce P-load cell with force resolution of 20 N and displacement resolution of ±2 μm. All tests were performed by displacing the upper grip at constant speed (2 mm min$^{-1}$ for AE experiments and optical microscopy; 25 mm min$^{-1}$ for IR microscopy), and the corresponding strain was calculated as the ratio of the displacement over the initial gauge length *L*.

AE sensors (micro-80 from EPAC, 10-mm diameter) were attached to the wider shoulders of samples SP1&2 on either side of the gauge section [regions R1 and R2 in Figure 1(a)], using a thin layer of Vaseline for improved acoustic coupling. The frequency response of the sensors was relatively broad and spanned 200 kHz – 1 MHz. For our AE studies, the signal from the sensors was first pre-amplified ($\times 10^3$) and then fed into the data-acquisition system (PCI-2 from EPAC). Individual AE events were identified and recorded when the pre-amplified input signal from any of the sensors exceeded the measurement threshold of 22 dB. The duration *D* of each AE event was determined as the time passed between when the signal first exceeded the measurement threshold, and the signal remained below the measurement threshold for 200 μs. The energy *E* of each AE event was determined as the ratio of the integral of the squared AE signal over the entire duration of the event *D*, and a reference resistance of 10 kΩ. AE activity was determined as the number of events per second, computed from the AE signal using constant intervals of one second each.

The location $z$ of AE events along the length of the gauge region was determined via $z = (L/2)(1 - \Delta t/\Delta t_{max})$, where $\Delta t$ is the time difference between the detection of the same event by each of the two AE sensors, and $\Delta t_{max} = 40$ μs is the time required for sound to travel between the two sensors when they are $L = 80$ mm apart, which was experimentally determined by emitting sound waves using one of the detectors, and recording them using the other detector. (Note that $\Delta t_{max}$ should be < 40 μs according to $L = 80$ mm and the speed of sound $v \sim 5000$ m s$^{-1}$ for polycrystalline Ni-Ti [53], but in practice the imperfect acoustic contact between sample and detectors introduced delays; note also that this formalism ignores small delays due to variable gauge length, and finite gauge width, which lead to a combined uncertainty of ~7% in location values along the length of the gauge region.) For the small number of AE events that were successfully located, $E$ was obtained as the geometric mean of the energy of the corresponding individual events, in order to correct for acoustic attenuation [54].

X-ray diffraction was performed using Bragg-Brentano geometry and Cu K$\alpha_1$ = 1.5406 Å in an X'pert PRO PANalytical PW3040 diffractometer, in the mechanically unloaded sample SP3, at room temperature.

Grayscale optical images were recorded for sample SP1 at 25 Hz using a Zwick/Roell video camera equipped with a 768×576-pixel 17.5-bit-resolution detector. IR images were recorded for sample SP4 at 115.5 Hz using an Infratech ImageIR MCT 8800 camera with a 640×512-pixel 8.0-10.2-μm-spectrum detector. The use of a microscope lens with 50-mm fixed focus led to a spatial resolution of ~34 μm per pixel. In order to record true temperature (with 0.04 K resolution), the sample was sprayed with a thin layer of matt black paint (PNM, electrolube) of known emissivity. Average sample surface temperature was obtained as the arithmetic mean of the temperature recorded at each pixel.

## III. RESULTS AND DISCUSSION

Figure 1(b) shows the recorded stress-AE activity versus time data for sample SP1. The AE activity during the first loading is high, but decreases dramatically during unloading, and remains low on further cycling until the 65$^{th}$ loading when the sample breaks down and there is a sudden increase in the AE activity due to fracture. Figure 1(c) shows the corresponding evolution with mechanical cycling of the stress-strain curves. For each cycle, the initial elastic behaviour on loading is followed by a stress plateau that is associated with the strain-driven B2-B19' martensitic transition (cf. Refs [2,48–50]; Supplementary Note 3, [52]). On unloading, strain is not reversibly recovered for the first 40 cycles, after which reversible superelasticity is achieved. This irreversibility in strain is understood to arise due to slip plastic deformations in the sample [31–33], whose resulting internal stresses lead to retention of martensite in the unloaded state, and work hardening [31–33]. Retained martensite favours the formation of strain-driven martensite on further cycling, thus reducing the value of the corresponding transformation stress [Figure 1(c)], and work-hardening results in plateaus with finite slope [Figure 1(c)]. We note that the large apparent irreversible strain after the first mechanical cycle arises primarily due to the sliding of the dog-bone-shaped sample at the grips of the testing machine during the initial stages of loading, suggesting that the corresponding high AE recorded during the first loading is mostly due to dynamic friction between the grips and the sample, as shown in more detail later. In order to confirm the presence of retained martensite, we performed x-ray diffraction in sample SP3 after the 1$^{st}$ and the 10$^{th}$ loading/unloading cycles (after #1 and after #10, Figure 2), and compared the resulting patterns with that obtained prior to mechanical testing (virgin, Figure 2). After cycle #1, there is no apparent increase in the amount of retained martensite. After 10 cycles, the small decrease in intensity in the austenite diffraction peak 110, and the simultaneous small increase in intensity in some of the martensite diffraction peaks, indicate that the amount of retained martensite increased somewhat with mechanical cycling. Retained martensite tends to form near slip plastic deformations, and its built-in internal stress tends to assist with the formation of strain-driven martensite [55], consistent with our observation of reduced transformation stress on mechanical cycling [Figure 1(c)].

Figure 3(a) shows details for a small number of selected loading/unloading cycles, namely initial cycles #1-3, and final cycles #64-65. For cycle #1, there is an intense AE activity on elastic loading (Figure 3(a), region 1), which is primarily due to dynamic friction between the grips of the mechanical testing machine and the sample, and plastic deformation induced by the grips, given that almost no AE events are detected within the gauge section as shown later. This extrinsic contribution to the AE, which we will not consider any longer, is followed by weaker AE activity during the strain-driven transition plateau on loading (Figure 3(a), region 2), and even weaker AE activity during unloading (Figure 3(a), region 3). Given that there is no apparent increase in retained martensite after the first mechanical cycle (Figure 2), we infer that the stronger AE recorded during the first loading arises primarily due to the changes in grain configuration and dislocation density associated with the nucleation and propagation of localised wide Lüders-like bands. For cycles #2-3, the AE activity is qualitatively similar but one order of magnitude less. We note that the degree of attenuation of ultrasound pulses travelling between the two AE sensors did not change between cycles #1 and #5, indicating that the observed suppression of AE activity is not due to increased acoustic losses. The last two cycles #64-65 display very low AE activity, apart from the intense AE activity due to sample fracture at the very end of the experiment. This latter AE activity arises within < 1s from mechanical failure, in agreement with a previous observation that Ni-Ti breakdown occurs shortly after crack nucleation [50].

For the same selected cycles, Figure 3(b) shows the spatial distribution of the AE events that were successfully located during the strain-driven transition on loading. For cycle #1, the located AE events during the transition plateau are spread across the gauge section. We note that only three AE events could be located within the gauge section during the first elastic loading, confirming that the corresponding intense AE activity [orange line in cycle #1, Figure 3(a)] arises almost exclusively due to grips-sample interactions. For cycle #2, the located AE events are concentrated in the bottom half of the gauge section, and for cycle #3, only a small number of AE events could be located due to the weak AE activity. For cycle #65, located AE events concentrate near the top shoulder of the dog-bone-shaped sample where fracture occurred, indicating that the location method is robust.

Figure 4 shows energy distributions of AE events, and of their duration, for cycles #1-65, during the strain-driven transition on loading [(a-d) and (g,h)] and unloading [(e,f)]. For cycle #1, the energy distribution of AE events on loading [Figure 4(a)] displays a power-law behaviour [37,56,57] with a small negative slope, indicating that most AE events are energetic and short. The corresponding time distribution of AE energies during the transition plateau [Figure 4(b)] displays a power-law behaviour with $D \sim E^{1/3}$ indicating that the transition dynamics is scale invariant. Cycles #2-64 display similar behaviour on loading [Figure 4(c,d)] and unloading [Figure 4(e,f)], but with poorer statistics due the small number of AE events recorded. Sample fracture on cycle #65 produced a small number of long energetic AE events [Figure 4(g,h)].

Figure 5 shows optical images recorded during the mechanical cycling described above, at strain value of 3%, which falls within the transition plateau. A small number of wide

Lüders-like bands are present during the first loading/unloading cycle, and the bands become more in number and gradually narrower on successive loading/unloading cycles, in agreement with previous observations [23]. The angle between the bands and the vertical loading axis is ~55°, which minimizes the strain incompatibility at the interface between the elastically strained austenite and the strain-induced martensite [58,59].

Figure 6 shows corresponding IR images on a separate sample SP4 that was cycled at a faster rate in order to detect thermal changes. Lüders-like bands are visible in the first loading/unloading cycle, but fade rapidly with further cycling [7]. The forward martensitic transition on loading results on heating of the sample, and the reverse transition on unloading results on cooling of the sample. The average temperature change $\Delta T$ on loading is ~26 K, and on unloading is ~ -28 K, which represent a giant elastocaloric effect [5–8]. Higher temperature leads to higher critical transition stress [7] and results in transition plateaus of finite slope [Figure 6(c)].

It is well known that martensitic transition fronts in Ni-Ti polycrystalline alloys under tension propagate across grains in a localised manner [7,23–27], in order to satisfy the strain compatibility between austenite and martensite on a macroscopic length scale. This macroscopic compatibility leads to enhanced local stresses in grains within the transition front, which permit reaching simultaneously critical transition stresses in grains of different orientation. The enhanced local stresses in transforming grains also lead to enhanced dislocation motion and plastic deformation via slip mechanisms [31], which can generate significant AE, and result in persistent changes in intergranular interactions via e.g. changes in relative orientation and size of grains, as observed using spatially resolved electron backscatter diffraction in Ref. [23].

Figure 7 shows time-dependent stress and AE activity plots for a separate sample SP2 that was subjected to two loading/unloading cycles (i) prior to thermal reset, (ii) after *ex-situ* 1-hour annealing at 100°C, and (iii) after *ex-situ* 1-hour annealing at 250°C. The strong AE activity recorded during the first transition plateau prior to annealing is not recovered after either thermal treatment, demonstrating that this activity arises primarily due to changes in grain configuration and dislocation density [31]. This is because annealing temperatures are too low, and annealing times are too short, to cause any significant recrystallization [60], such that the AE activity recorded after either thermal treatment arises primarily due to the martensitic transition in the accommodated grains. (The intense AE activity that arises during each of the initial elastic loadings is again due to sliding of the dog-bone-shaped sample at the grips that occurs after each mounting of the sample in the mechanical testing machine.)

Rather unusually, after the first loading the AE that arises from the martensitic transition is weak. We attribute this to the good elastic compatibility between the B2 austenitic phase and the B19' martensitic phase [61,62], and to the very low value of the elastic constants associated with the two deformation modes that underpin this transition [63]. These two factors may result in a suppression of the elastic energy associated with differences in the lattice of the two transforming phases, and therefore in weak AE during the transition within and across the accommodated grains. It has been proposed that the B2 cubic phase transforms into the B19' monoclinic phase via the combination of shear and shuffle strains of basal {110} planes along <1-10> directions, and shear strains of non-basal {001} planes along <1-10> directions [3]. The first deformation mode is associated with the shear elastic constant $c' = (c_{11}-c_{12})/2$, which is usually low for alloys that display martensitic transitions ($< 20$ GPa) [18,63–70]. The second deformation mode is associated with the shear elastic constant $c_{44}$, which is usually high for alloys that display martensitic transitions (~100 GPa) [65–70], yielding large values of elastic anisotropy $A = c_{44}/c' = 2c_{44}/(c_{11}-c_{12})$. ($c_{11}$, $c_{12}$ and $c_{44}$ are the three independent elastic constants for cubic systems.) The unusually low values of $c_{44}$ for Ni-Ti alloys (< 40 GPa) [18,63,64] lead to unusually low values of elastic anisotropy $A$ for the B2 austenitic phase, particularly near the martensitic transition where both shear moduli soften [63]. For example, $A \sim 2$ for Ni-Ti [63,64], a value that is 7 times lower than those for Cu-based shape memory alloys [65–67], and 3-5 times lower than those for Ni-Mn-based shape memory alloys [68–70], for both of which only one soft deformation mode is available [shear and shuffle strains of basal {110} planes along <1-10> directions], and display strong AE activity when driving the transition mechanically (Supplementary Note 4 in [52] and Refs [34,43,44]) or thermally [57,71,72]. Our experimental observation of weak AE in mechanically cycled Ni-rich Ni-Ti polycrystalline alloys is consistent with recent phase-field models [73,74] that predict that martensitic transitions in alloys with low elastic anisotropy become smoother and take place via low-energy avalanches, which are challenging to detect using AE. On sustained mechanical cycling, the weak AE becomes even weaker due to accumulation of retained martensite, which favours the formation of strain-driven martensite and weakens the first-order character of the B2-B19' transition [31–33].

## IV. SUMMARY AND CONCLUSIONS

We investigated changes in AE on mechanically cycling polycrystalline $Ni_{50.4}Ti_{49.6}$ alloys across the tensile-strain-driven B2-B19' structural phase transition that is at the core of their superelastic/elastocaloric properties. Persistent changes in the configuration of grains that are forced by elastic compatibility between transforming grains of different orientation lead to strong AE activity only during the first mechanical loading.

The AE activity associated with the martensitic transition in the accommodated polycrystal remains weak throughout cycling, because the elastic compatibility between the transforming phases is high, the elastic anisotropy of Ni-Ti is low, and the first-order character of the transition weakens with increased retained martensite.

## ACKNOWLEDGEMENTS

EV, AP, and MR acknowledge financial support from the Spanish Ministry of Science, Project MAT2016-75823-R. GFN thanks the Royal Commission for the Exhibition of 1851 for the award of a Research Fellowship. XM is grateful for support from the Royal Society.

# REFERENCES

[1] M.H. Elahinia, M. Hashemi, M. Tabesh, S.B. Bhaduri, Manufacturing and processing of NiTi implants: A review, Prog. Mater. Sci. 57 (2012) 911–946. doi:10.1016/j.pmatsci.2011.11.001.

[2] K. Otsuka, X. Ren, Physical metallurgy of Ti–Ni-based shape memory alloys, Prog. Mater. Sci. 50 (2005) 511–678. doi:10.1016/j.pmatsci.2004.10.001.

[3] K. Otsuka, C.M. Wayman, eds., Shape Memory Materials, Cambridge, 1999.

[4] J. Mohd Jani, M. Leary, A. Subic, M.A. Gibson, A review of shape memory alloy research, applications and opportunities, Mater. Des. 56 (2014) 1078–1113. doi:10.1016/j.matdes.2013.11.084.

[5] J. Cui, Y. Wu, J. Muehlbauer, Y. Hwang, R. Radermacher, S. Fackler, M. Wuttig, I. Takeuchi, Demonstration of high efficiency elastocaloric cooling with large $\Delta T$ using NiTi wires, Appl. Phys. Lett. 101 (2012) 073904. doi:10.1063/1.4746257.

[6] J. Tušek, K. Engelbrecht, L.P. Mikkelsen, N. Pryds, Elastocaloric effect of Ni-Ti wire for application in a cooling device, J. Appl. Phys. 117 (2015) 124901. doi:10.1063/1.4913878.

[7] H. Ossmer, C. Chluba, M. Gueltig, E. Quandt, M. Kohl, Local Evolution of the Elastocaloric Effect in TiNi-Based Films, Shape Mem. Superelasticity. 1 (2015) 142–152. doi:10.1007/s40830-015-0014-3.

[8] X. Moya, S. Kar-Narayan, N.D. Mathur, Caloric materials near ferroic phase transitions, Nat. Mater. 13 (2014) 439–450. doi:10.1038/nmat3951.

[9] M. Schmidt, A. Schütze, S. Seelecke, Scientific test setup for investigation of shape memory alloy based elastocaloric cooling processes, Int. J. Refrig. 54 (2015) 88–97. doi:10.1016/j.ijrefrig.2015.03.001.

[10] F. Welsch, S.-M. Kirsch, N. Michaelis, P. Motzki, M. Schmidt, A. Schütze, S. Seelecke, Elastocaloric Cooling: System Design, Simulation, and Realization, in: Vol. 2 Mech. Behav. Act. Mater. Struct. Heal. Monit. Bioinspired Smart Mater. Syst. Energy Harvest. Emerg. Technol., American Society of Mechanical Engineers, 2018. doi:10.1115/SMASIS2018-7982.

[11] M. Schmidt, A. Schütze, S. Seelecke, Elastocaloric cooling processes: The influence of material strain and strain rate on efficiency and temperature span, APL Mater. 4 (2016) 064107. doi:10.1063/1.4953433.

[12] H. Ossmer, C. Chluba, S. Kauffmann-Weiss, E. Quandt, M. Kohl, TiNi-based films for elastocaloric microcooling— Fatigue life and device performance, APL Mater. 4 (2016) 064102. doi:10.1063/1.4948271.

[13] F. Bruederlin, H. Ossmer, F. Wendler, S. Miyazaki, M. Kohl, SMA foil-based elastocaloric cooling: from material behavior to device engineering, J. Phys. D. Appl. Phys. 50 (2017) 424003. doi:10.1088/1361-6463/aa87a2.

[14] S. Qian, Y. Geng, Y. Wang, J. Muehlbauer, J. Ling, Y. Hwang, R. Radermacher, I. Takeuchi, Design of a hydraulically driven compressive elastocaloric cooling system, Sci. Technol. Built Environ. 22 (2016) 500–506. doi:10.1080/23744731.2016.1171630.

[15] H. Hou, J. Cui, S. Qian, D. Catalini, Y. Hwang, R. Radermacher, I. Takeuchi, Overcoming fatigue through compression for advanced elastocaloric cooling, MRS Bull. 43 (2018) 285–290. doi:10.1557/mrs.2018.70.

[16] J. Tušek, K. Engelbrecht, D. Eriksen, S. Dall'Olio, J. Tušek, N. Pryds, A regenerative elastocaloric heat pump, Nat. Energy. 1 (2016) 16134.

doi:10.1038/nenergy.2016.134.

[17] S. Qian, A. Alabdulkarem, J. Ling, J. Muehlbauer, Y. Hwang, R. Radermacher, I. Takeuchi, Performance enhancement of a compressive thermoelastic cooling system using multi-objective optimization and novel designs, Int. J. Refrig. 57 (2015) 62–76. doi:10.1016/j.ijrefrig.2015.04.012.

[18] O. Mercier, K.N. Melton, G. Gremaud, J. Hägi, Single-crystal elastic constants of the equiatomic NiTi alloy near the martensitic transformation, J. Appl. Phys. 51 (1980) 1833–1834. doi:10.1063/1.327750.

[19] K. Gall, H. Sehitoglu, The role of texture in tension–compression asymmetry in polycrystalline NiTi, Int. J. Plast. 15 (1999) 69–92. doi:10.1016/S0749-6419(98)00060-6.

[20] K. Gall, H. Sehitoglu, R. Anderson, I. Karaman, Y.I. Chumlyakov, I. V. Kireeva, On the mechanical behavior of single crystal NiTi shape memory alloys and related polycrystalline phenomenon, Mater. Sci. Eng. A. 317 (2001) 85–92. doi:10.1016/S0921-5093(01)01183-2.

[21] K. Gall, J. Tyber, G. Wilkesanders, S.W. Robertson, R.O. Ritchie, H.J. Maier, Effect of microstructure on the fatigue of hot-rolled and cold-drawn NiTi shape memory alloys, Mater. Sci. Eng. A. 486 (2008) 389–403. doi:10.1016/j.msea.2007.11.033.

[22] F.M. Weafer, Y. Guo, M.S. Bruzzi, The effect of crystallographic texture on stress-induced martensitic transformation in NiTi: A computational analysis, J. Mech. Behav. Biomed. Mater. 53 (2016) 210–217. doi:10.1016/j.jmbbm.2015.08.023.

[23] L. Wang, L. Ma, C. Liu, Z.Y. Zhong, S.N. Luo, Texture-induced anisotropic phase transformation in a NiTi shape memory alloy, Mater. Sci. Eng. A. 718 (2018) 96–103. doi:10.1016/j.msea.2018.01.075.

[24] X. Bian, A.A. Saleh, E. V. Pereloma, C.H.J. Davies, A.A. Gazder, A digital image correlation study of a NiTi alloy subjected to monotonic uniaxial and cyclic loading-unloading in tension, Mater. Sci. Eng. A. 726 (2018) 102–112. doi:10.1016/j.msea.2018.04.081.

[25] S.C. Mao, J.F. Luo, Z. Zhang, M.H. Wu, Y. Liu, X.D. Han, EBSD studies of the stress-induced B2–B19′ martensitic transformation in NiTi tubes under uniaxial tension and compression, Acta Mater. 58 (2010) 3357–3366. doi:10.1016/j.actamat.2010.02.009.

[26] P. Sedmák, J. Pilch, L. Heller, J. Kopeček, J. Wright, P. Sedlák, M. Frost, P. Šittner, Grain-resolved analysis of localized deformation in nickel-titanium wire under tensile load, Science (80-. ). 353 (2016) 559–562. doi:10.1126/science.aad6700.

[27] E.A. Pieczyska, S.P. Gadaj, W.K. Nowacki, H. Tobushi, Phase-Transformation Fronts Evolution for Stress- and Strain-Controlled Tension Tests in TiNi Shape Memory Alloy, Exp. Mech. 46 (2006) 531–542. doi:10.1007/s11340-006-8351-y.

[28] P. Šittner, Y. Liu, V. Novak, On the origin of Lüders-like deformation of NiTi shape memory alloys, J. Mech. Phys. Solids. 53 (2005) 1719–1746. doi:10.1016/j.jmps.2005.03.005.

[29] M. Frost, P. Sedlák, T. Ben Zineb, Experimental Observations and Modeling of Localization in Superelastic NiTi Polycrystalline Alloys: State of the Art, Acta Phys. Pol. A. 134 (2018) 847–852. doi:10.12693/APhysPolA.134.847.

[30] L. Zheng, Y. He, Z. Moumni, Lüders-like band front motion and fatigue life of pseudoelastic polycrystalline NiTi shape memory alloy, Scr. Mater. 123 (2016) 46–50. doi:10.1016/j.scriptamat.2016.05.042.

[31] P. Sedmák, P. Šittner, J. Pilch, C. Curfs, Instability of cyclic superelastic deformation of NiTi investigated by synchrotron X-ray diffraction, Acta Mater. 94 (2015) 257–270. doi:10.1016/j.actamat.2015.04.039.

[32] P. Chowdhury, H. Sehitoglu, A revisit to atomistic rationale for slip in shape memory alloys, Prog. Mater. Sci. 85 (2017) 1–42. doi:10.1016/j.pmatsci.2016.10.002.

[33] P. Šittner, P. Sedlák, H. Seiner, P. Sedmák, J. Pilch, R. Delville, L. Heller, L. Kaděřávek, On the coupling between martensitic transformation and plasticity in NiTi: Experiments and continuum based modelling, Prog. Mater. Sci. 98 (2018) 249–298. doi:10.1016/j.pmatsci.2018.07.003.

[34] L. Straka, V. Novák, M. Landa, O. Heczko, Acoustic emission of Ni–Mn–Ga magnetic shape memory alloy in different straining modes, Mater. Sci. Eng. A. 374 (2004) 263–269. doi:10.1016/j.msea.2004.03.018.

[35] S. Sreekala, G. Ananthakrishna, Acoustic Emission and Shape Memory Effect in the Martensitic Transformation, Phys. Rev. Lett. 90 (2003) 135501. doi:10.1103/PhysRevLett.90.135501.

[36] A. Amengual, F. Garcias, F. Marco, C. Segui, V. Torra, Acoustic emission of the interface motion in the martensitic transformation of Cu-Zn-Al shape memory alloy, Acta Metall. 36 (1988) 2329–2334. doi:10.1016/0001-6160(88)90332-X.

[37] A. Planes, L. Mañosa, E. Vives, Acoustic emission in martensitic transformations, J. Alloys Compd. 577 (2013) S699–S704. doi:10.1016/j.jallcom.2011.10.082.

[38] A. Vinogradov, I.S. Yasnikov, On the nature of acoustic emission and internal friction during cyclic deformation of metals, Acta Mater. 70 (2014) 8–18. doi:10.1016/j.actamat.2014.02.007.

[39] M. Shaira, N. Godin, P. Guy, L. Vanel, J. Courbon, Evaluation of the strain-induced martensitic transformation by acoustic emission monitoring in 304L austenitic stainless steel: Identification of the AE signature of the martensitic transformation and power-law statistics, Mater. Sci. Eng. A. 492 (2008) 392–399. doi:10.1016/j.msea.2008.04.068.

[40] C.K. Mukhopadhyay, K. V. Kasiviswanathan, T. Jayakumar, B. Raj, Acoustic emission during tensile deformation of annealed and cold-worked AISI type 304 austenitic stainless steel, J. Mater. Sci. 28 (1993) 145–154. doi:10.1007/BF00349045.

[41] S.H. Carpenter, F.P. Higgins, Sources of acoustic emission generated during the plastic deformation of 7075 aluminum alloy, Metall. Trans. A. 8 (1977) 1629–1632. doi:10.1007/BF02644869.

[42] M.G.R. Sause, In Situ Monitoring of Fiber-Reinforced Composites, Springer International Publishing, 2016. doi:10.1007/978-3-319-30954-5.

[43] E. Vives, D. Soto-Parra, L. Mañosa, R. Romero, A. Planes, Driving-induced crossover in the avalanche criticality of martensitic transitions, Phys. Rev. B. 80 (2009) 180101. doi:10.1103/PhysRevB.80.180101.

[44] E. Bonnot, E. Vives, L. Mañosa, A. Planes, R. Romero, Acoustic emission and energy dissipation during front propagation in a stress-driven martensitic transition, Phys. Rev. B. 78 (2008) 094104. doi:10.1103/PhysRevB.78.094104.

[45] C. Dunand-Châtellet, Z. Moumni, Experimental analysis of the fatigue of shape memory alloys through power-law statistics, Int. J. Fatigue. 36 (2012) 163–170. doi:10.1016/j.ijfatigue.2011.07.014.

[46] E.A. Pieczyska, H. Tobushi, K. Takeda, D. Stróż, Z. Ranachowski, K. Kulasiński, S. Kúdela Jr., J. Luckner, Martensite transformation bands studied in

TiNi shape memory alloy by infrared and acoustic emission techniques, Met. Mater. 50 (2012) 309–318. doi:10.4149/km_2012_5_309.

[47] A.C. Lucia, C. Santulli, Study of Mechanical Behaviour in Shape Memory Alloys Through Adiabatic Thermal Emission and Acoustic Emission Technique, J. Phys. IV France 07 (1997) C5-637-C5-642. doi:10.1051/jp4:19975101.

[48] G. Laplanche, T. Birk, S. Schneider, J. Frenzel, G. Eggeler, Effect of temperature and texture on the reorientation of martensite variants in NiTi shape memory alloys, Acta Mater. 127 (2017) 143–152. doi:10.1016/j.actamat.2017.01.023.

[49] J. Tušek, A. Žerovnik, M. Čebron, M. Brojan, B. Žužek, K. Engelbrecht, A. Cadelli, Elastocaloric effect vs fatigue life: Exploring the durability limits of Ni-Ti plates under pre-strain conditions for elastocaloric cooling, Acta Mater. 150 (2018) 295–307. doi:10.1016/j.actamat.2018.03.032.

[50] S.W. Robertson, A.R. Pelton, R.O. Ritchie, Mechanical fatigue and fracture of Nitinol, Int. Mater. Rev. 57 (2012) 1–37. doi:10.1179/1743280411Y.0000000009.

[51] E. Bonnot, R. Romero, X. Illa, L. Mañosa, A. Planes, E. Vives, Hysteresis in a system driven by either generalized force or displacement variables: Martensitic phase transition in single-crystalline Cu-Zn-Al, Phys. Rev. B. 76 (2007) 064105. doi:10.1103/PhysRevB.76.064105.

[52] See Supplemental Material at [URL will be inserted by publisher] for details on thermally driven transitions, lateral surface roughness, strain-driven B2-B19' transition, and AE recorded on mechanically cycling a superelastic single crystal with high elastic anisotropy.

[53] D. Bradley, Sound Propagation in Near-Stoichiometric Ti-Ni Alloys, J. Acoust. Soc. Am. 37 (1965) 700–702. doi:10.1121/1.1909397.

[54] R. Niemann, J. Kopeček, O. Heczko, J. Romberg, L. Schultz, S. Fähler, E. Vives, L. Mañosa, A. Planes, Localizing sources of acoustic emission during the martensitic transformation, Phys. Rev. B. 89 (2014) 214118. doi:10.1103/PhysRevB.89.214118.

[55] S. Miyazaki, T. Imai, Y. Igo, K. Otsuka, Effect of cyclic deformation on the pseudoelasticity characteristics of Ti-Ni alloys, Metall. Trans. A. 17 (1986) 115–120. doi:10.1007/BF02644447.

[56] L. Carrillo, L. Mañosa, J. Ortín, A. Planes, E. Vives, Experimental Evidence for Universality of Acoustic Emission Avalanche Distributions during Structural Transitions, Phys. Rev. Lett. 81 (1998) 1889–1892. doi:10.1103/PhysRevLett.81.1889.

[57] M.-L. Rosinberg, E. Vives, Metastability, Hysteresis, Avalanches, and Acoustic Emission: Martensitic Transitions in Functional Materials, in: Springer Ser. Mater. Sci., 2012: pp. 249–272. doi:10.1007/978-3-642-20943-7_13.

[58] J.A. Shaw, Simulations of localized thermo-mechanical behavior in a NiTi shape memory alloy, Int. J. Plast. 16 (2000) 541–562. doi:10.1016/S0749-6419(99)00075-3.

[59] L. Dong, R.H. Zhou, X.L. Wang, G.K. Hu, Q.P. Sun, On interfacial energy of macroscopic domains in polycrystalline NiTi shape memory alloys, Int. J. Solids Struct. 80 (2016) 445–455. doi:10.1016/j.ijsolstr.2015.10.006.

[60] Y. Liu, J. Van Humbeeck, R. Stalmans, L. Delaey, Some aspects of the properties of NiTi shape memory alloy, J. Alloys Compd. 247 (1997) 115–121. doi:10.1016/S0925-8388(96)02572-8.

[61] S. Dilibal, H. Sehitoglu, R.F. Hamilton, H.J. Maier, Y. Chumlyakov, On the volume change in Co–Ni–Al during pseudoelasticity, Mater. Sci. Eng. A. 528

[61] (2011) 2875–2881. doi:10.1016/j.msea.2010.12.056.

[62] A. Ahadi, Q. Sun, Stress-induced nanoscale phase transition in superelastic NiTi by in situ X-ray diffraction, Acta Mater. 90 (2015) 272–281. doi:10.1016/j.actamat.2015.02.024.

[63] X. Ren, K. Otsuka, The Role of Softening in Elastic Constant $c_{44}$ in Martensitic Transformation, Scr. Mater. 38 (1998) 1669–1675. doi:10.1016/S1359-6462(98)00078-5.

[64] T.M. Brill, S. Mittelbach, W. Assmus, M. Mullner, B. Luthi, Elastic properties of NiTi, J. Phys. Condens. Matter. 3 (1991) 9621–9627. doi:10.1088/0953-8984/3/48/004.

[65] G. Guenin, M. Morin, P.F. Gobin, W. Dejonghe, L. Delaey, Elastic constant measurements in β Cu-Zn-Al near the martensitic transformation temperature, Scr. Metall. 11 (1977) 1071–1075. doi:10.1016/0036-9748(77)90310-6.

[66] L. Mañosa, M. Jurado, A. Planes, J. Zarestky, T. Lograsso, C. Stassis, Elastic constants of bcc Cu-Al-Ni alloys, Phys. Rev. B. 49 (1994) 9969–9972. doi:10.1103/PhysRevB.49.9969.

[67] P. Sedlák, H. Seiner, M. Landa, V. Novák, P. Šittner, L. Mañosa, Elastic constants of bcc austenite and 2H orthorhombic martensite in CuAlNi shape memory alloy, Acta Mater. 53 (2005) 3643–3661. doi:10.1016/j.actamat.2005.04.013.

[68] M. Stipcich, L. Mañosa, A. Planes, M. Morin, J. Zarestky, T. Lograsso, C. Stassis, Elastic constants of Ni−Mn−Ga magnetic shape memory alloys, Phys. Rev. B. 70 (2004) 054115. doi:10.1103/PhysRevB.70.054115.

[69] X. Moya, L. Mañosa, A. Planes, T. Krenke, M. Acet, M. Morin, J.L. Zarestky, T.A. Lograsso, Temperature and magnetic-field dependence of the elastic constants of Ni-Mn-Al magnetic Heusler alloys, Phys. Rev. B. 74 (2006) 024109. doi:10.1103/PhysRevB.74.024109.

[70] X. Moya, D. González-Alonso, L. Mañosa, A. Planes, V.O. Garlea, T.A. Lograsso, D.L. Schlagel, J.L. Zarestky, S. Aksoy, M. Acet, Lattice dynamics in magnetic superelastic Ni-Mn-In alloys: Neutron scattering and ultrasonic experiments, Phys. Rev. B. 79 (2009) 214118. doi:10.1103/PhysRevB.79.214118.

[71] M.C. Gallardo, J. Manchado, F.J. Romero, J. del Cerro, E.K.H. Salje, A. Planes, E. Vives, R. Romero, M. Stipcich, Avalanche criticality in the martensitic transition of $Cu_{67.64}Zn_{16.71}Al_{15.65}$ shape-memory alloy: A calorimetric and acoustic emission study, Phys. Rev. B. 81 (2010) 174102. doi:10.1103/PhysRevB.81.174102.

[72] B. Ludwig, C. Strothkaemper, U. Klemradt, X. Moya, L. Mañosa, E. Vives, A. Planes, An acoustic emission study of the effect of a magnetic field on the martensitic transition in $Ni_2MnGa$, Appl. Phys. Lett. 94 (2009) 121901. doi:10.1063/1.3103289.

[73] M. Porta, T. Castán, P. Lloveras, A. Saxena, A. Planes, Intermittent dynamics in externally driven ferroelastics and strain glasses, Phys. Rev. E. 98 (2018) 032143. doi:10.1103/PhysRevE.98.032143.

[74] P. Lloveras, T. Castán, M. Porta, A. Planes, A. Saxena, Influence of Elastic Anisotropy on Structural Nanoscale Textures, Phys. Rev. Lett. 100 (2008) 165707. doi:10.1103/PhysRevLett.100.165707.

[75] H. Sitepu, Texture and structural refinement using neutron diffraction data from molybdite ($MoO_3$) and calcite ($CaCO_3$) powders and a Ni-rich $Ni_{50.7}Ti_{49.30}$ alloy, Powder Diffr. 24 (2009) 315–326. doi:10.1154/1.3257906.

[76] Y. Kudoh, M. Tokonami, S. Miyazaki, K. Otsuka, Crystal structure of the martensite in Ti-49.2 at.%Ni alloy analyzed by the single crystal X-ray diffraction method, Acta Metall. 33 (1985) 2049–2056. doi:10.1016/0001-6160(85)90128-2.

[77] K. Engelbrecht, J. Tušek, S. Sanna, D. Eriksen, O. V. Mishin, C.R.H. Bahl, N. Pryds, Effects of surface finish and mechanical training on Ni-Ti sheets for elastocaloric cooling, APL Mater. 4 (2016) 064110. doi:10.1063/1.4955131.

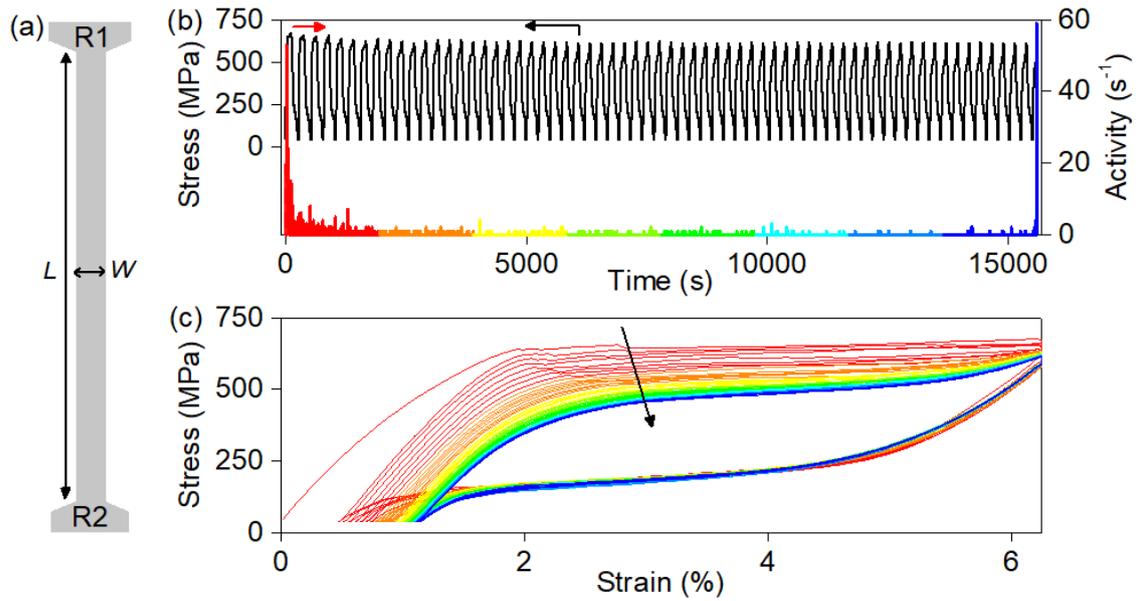

FIG. 1. AE on mechanically cycling the B2-B19' martensitic transition. (a) Schematic of the dog-bone shaped samples used for mechanical testing, with gauge length $L = 80$ mm and gauge width $W = 10$ mm. R1 and R2 indicate the position of the AE sensors. (b) Stress (left-hand-side axis) and AE activity (right-hand-side axis) as a function of time. (c) Corresponding stress-strain curves. AE activity data in (b) and stress-strain data in (c) are colour coded according to cycle number (–, cycles #1-7; –, cycles #8-15; –, cycles #16-23; –, cycles #24-32; –, cycles #33-40; –, cycles #41-48; –, cycles #49-56; –, cycles #57-65). Black arrow in (c) represents time. Data taken for sample SP1.

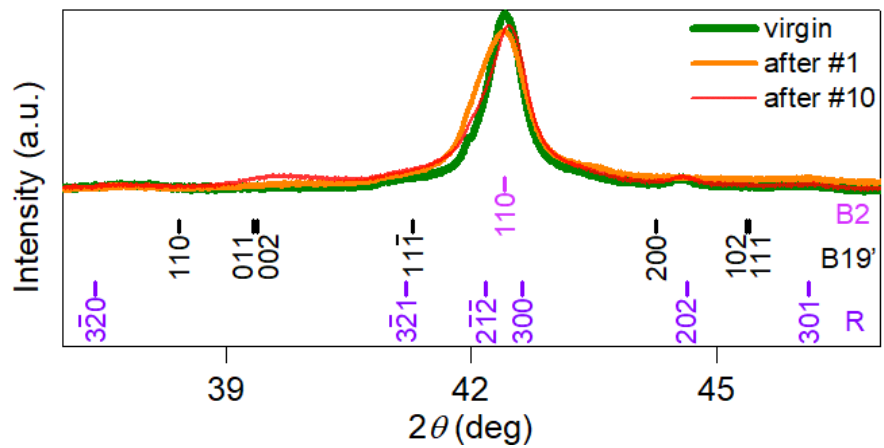

FIG. 2. Detail of selected x-ray diffraction spectra recorded before mechanical cycling, (virgin, −), after the first mechanical cycle (−, after #1), and after the tenth mechanical cycle (−, after #10). Vertical pink lines indicate indexed Bragg reflections for the austenitic B2 phase (Ref. [75]); vertical black lines indicate indexed reflections for the martensitic B19' phase (Ref. [76]); vertical purple lines indicate indexed reflections for the martensitic R phase (Ref. [75]). Data taken for sample SP3.

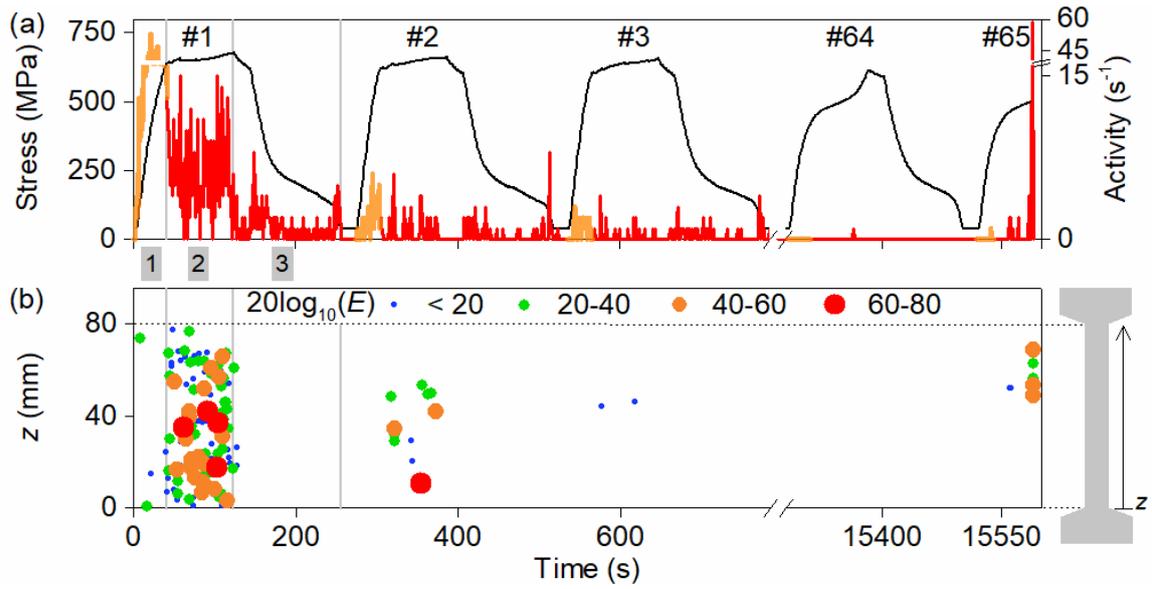

FIG. 3. Spatially resolved AE. For selected mechanical cycles (#1-3 and #64-65) in Figure 1(b), we plot (a) stress (left-hand-side axis) and AE activity (right-hand-side axis) as a function of time, and (b) location of individual AE events along $z$, grouped according to their energy $E$. In (a) and (b), grey vertical lines separate elastic loading (region 1), transition plateau on loading (region 2) and transition plateau on unloading (region 3). In (a), — represents AE activity recorded during elastic loading, and — represents AE activity recorded during the rest of the mechanical cycle. In (b), the size of the symbols represents their energy. Data taken for sample SP1.

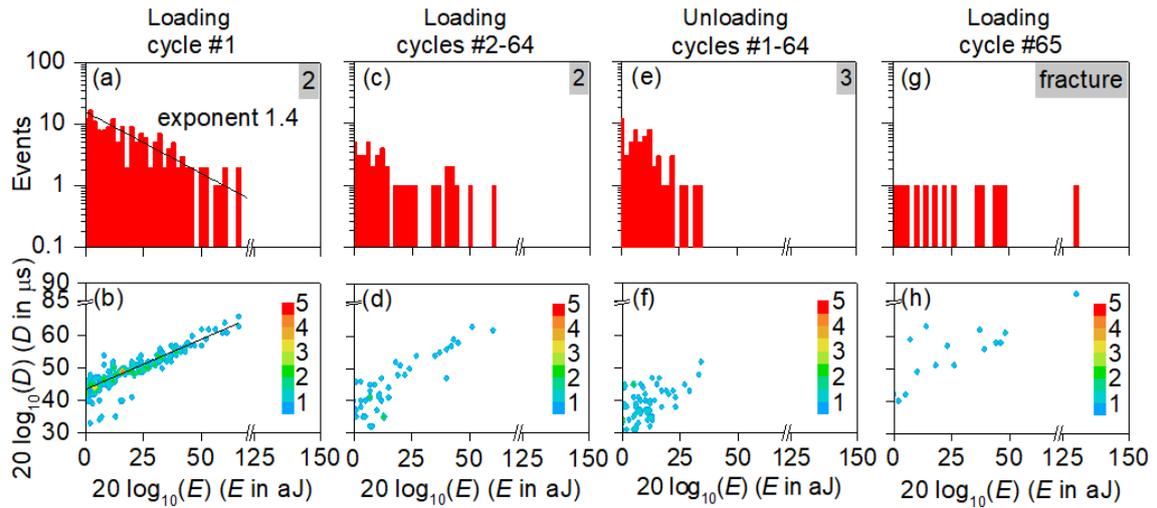

FIG. 4. Energy distributions. For the loading transition plateau during cycle #1 in Figure 3(a) (region 2), we plot the distribution in energy of (a) AE events and (b) duration. Panels (c-d) show corresponding data collected for subsequent cycles #2-64. For the unloading transition plateau [region 3 in Figure 3(a)], we plot the distribution in energy of (e) AE events and (f) duration for cycles #1-64. For the ultimate mechanical fracture that occurred during cycle #65, we plot the distribution in energy of (e) AE events and (f) duration. Black lines in (a-b) represent power-law fits, with exponents of 1.4 and 1/3, respectively. Colour scales in (b, d, f & h) represent number of AE events. Data taken for sample SP1.

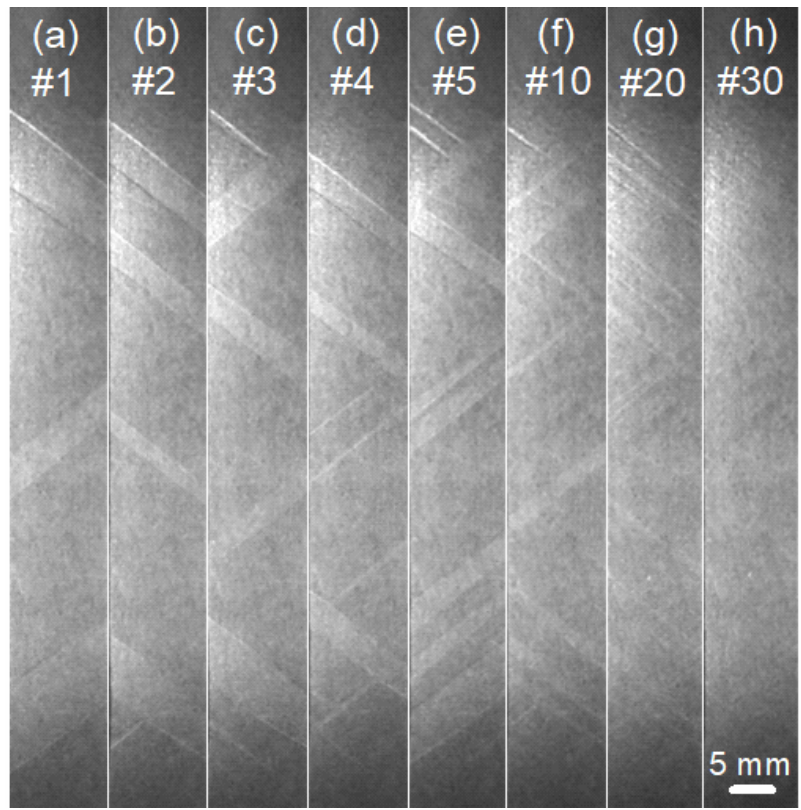

FIG. 5. Optical imaging of Lüders-like bands. Images of the gauge section, taken at strain values of 3% that fall within the transition plateau for cycles #1 (a), #2 (b), #3 (c), #4 (d), #5 (e), #10 (f), #20 (g), and #30 (h). For cycle #1, a complete set of images for the full mechanical loading is available via Supplementary Video S1. Data taken for sample SP1.

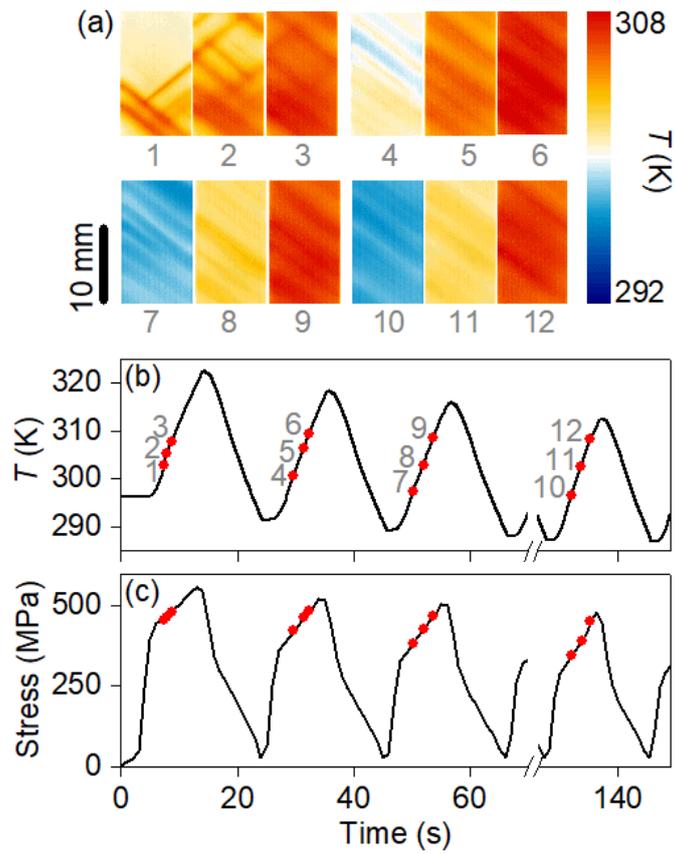

FIG. 6. IR imaging of Lüders-like bands. (a) Images of the central gauge section, taken during the transition plateau for cycles #1-3 and #7. (b) Average sample temperature as a function of time. The corresponding stress-time plots are shown in (c). Grey numbers and red symbols in panels (a-c) indicate the instants in time when the IR images were recorded. A complete set of images for the full mechanical loading is available via Supplementary Video S2. Data taken for sample SP4, which broke down after 200 cycles, in agreement with previous observations [49,77].

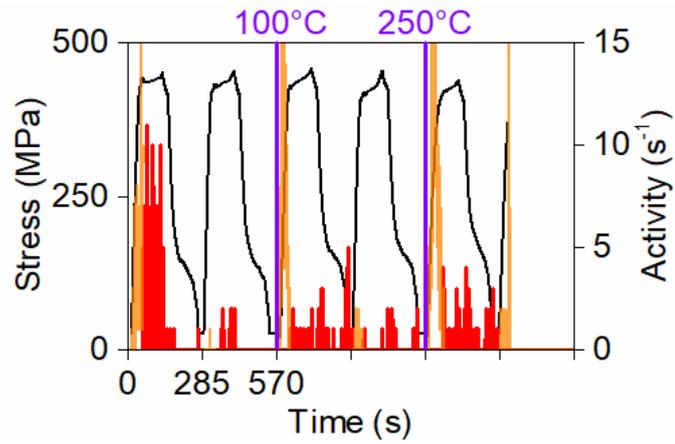

FIG. 7. Thermal reset. Stress (left-hand-side axis) and AE activity (right-hand-side axis) as a function of time, recorded prior to thermal reset, after *ex-situ* 1-hour annealing at 100°C, and after *ex-situ* 1-hour annealing at 250°C. Vertical purple lines indicate when the thermal treatments were performed. — represents AE activity recorded during elastic loading, and — represents AE activity recorded during the rest of the mechanical cycle. The sample broke down during the 6th mechanical cycle. Data taken for sample SP2.